\documentclass[twocolumn,showpacs,preprintnumbers,amsmath,amssymb]{revtex4}

\usepackage{graphicx}
\usepackage{dcolumn}
\usepackage{bm}

\begin{document}
\title{Collective diffusion of a colloidal particles in a liquid crystal. }
\author{B. I. Lev and A. G. Zagorodny}
\affiliation{Bogolyubov Institute for Theoretical Physics National Academy of Science of Ukraine,
Metrologichna Str. 14-b, 03680 Kyiv, Ukraine}
\date{\today}


\begin{abstract}
The collective diffusion effects in system of a colloidal particles in a liquid crystal
has been proposed. In this article described peculiarity of collective diffusion colloidal 
particles in a liquid crystal, which can be observe experimentally. The diffusion coefficient 
should be crucial dependence from temperature and concentration of particles. In system 
colloidal particles arise from the elastic distortion of the elastic director field the 
inter-particle interaction. These interaction can cause nontrivial collective behavior, 
as the results the non-monotonic dependence collective diffusion. Are predicted a nontrivial 
behavior collective diffusion of colloidal particles in a liquid crystal.
\end{abstract}
\maketitle

Colloidal particles in liquid crystals have attracted a great research interest during the last 
years. Anisotropic properties of the host fluid - liquid crystal give rise to a new class of a 
colloidal anisotropic interactions that never occurs in isotropic hosts. Liquid crystal colloidal 
system have much recent interest as models for diverse phenomena in condense matter physics. 
The anisotropic interactions are result an different structures of colloidal particles such as 
linear chains in inverter nematic emulsions \cite{po1,po2}, 2D crystals \cite{Mus} and 2D hexagonal 
structures at nematic-air interface \cite{nych}. The authors of \cite{nych1} observed 3D crystal 
structures in the system of hard particles with dipole configuration deformation of the director field. 

Recently \cite{tur} was observed the anomalous diffusion separate colloidal particle occurs at time 
scales that correspond to the relaxation times of director deformations around the particle. Once the 
nematic melts, the diffusion becomes normal and isotropic. Was shows that the deformations and 
fluctuations of the elastic director field profoundly influence diffusive regimes and was obtain as 
states are respectively termed sub-diffusion and super-diffusion. 

In this article will be investigate the collective behavior of the diffusion that the results 
particles interaction with formation new spatially nonuniform distribution \cite{lev}, \cite{lev2}, 
\cite{lev1}. Will be study the collective behavior the system colloidal particles. Nontrivial 
behavior collective diffusion of colloidal particles in liquid crystal has been predicted. 
Obtained results are in the temperature dependence of diffusion coefficient at various 
concentration, which have non monotonic character and transparent the new physical behavior 
the colloidal liquid crystal with taking into account the interaction between particles. 

For colloidal particles in solution, collective ͑also called cooperative or mutual͒ diffusion 
coefficient \cite{oh}- \cite{ver} is often determined experimentally with the help of dynamic 
light scattering. If one extrapolates this diffusion coefficient to a vanishing concentration 
of particles, it reduces to the single-particle diffusion coefficient since the interactions 
between the particles are presumably negligible then. At various concentration, particle 
interactions influence on the diffusion. At low enough concentrations, where many body 
interactions may be disregarded character of concentration will be play crucial role in 
the determination of the diffusion coefficient.

Complex investigation phase transition, collective diffusion, an auto-correlation function 
in this case allow the give the new physical picture collective behavior colloid media which 
a based on the liquid crystal. The critical non-monotonic engagement of both the temperature 
dependence viscus-elastic response functions is found to far more pronounces than for particle 
systems as a results of long-ranged interaction between the colloidal particles by induce the 
deformation elastic director field. The temperature and concentration dependence viscous elastic 
response of a suspension of spherical colloid particles in the vicinity of the isotropic- nematic 
is analyzed in the men-field region. Explicit for the share rate temperature and concentration 
dependence of the static structure factor are derived. Macroscopic expression for the anomalous 
part of the collective diffusion coefficient are derived, which are then expression as behavior 
the one the particle with the static structure factor with the elastic deformation director field. 
The temperature dependence of the drag force in system colloidal particles are motivation the 
nontrivial behavior collective diffusion of colloidal particles in liquid crystal.

Colloidal liquid crystal are characteristically systems with structure and time scales such 
that typical shear rate can bring them out of equilibrium and into some exotic states. Here 
we focus our attention on the diffusion motion the particle which are format this state. 
The rheological properties of liquid crystal and the structure of phase-separating (size, 
shape of patterns and their spatial distribution ) strongly affect the physical properties 
of the resulting colloids. These aspects have been extensive studied recently, and there has 
been a considerable progress in our understanding of the relation between structures and 
rheological properties. It is expected that the breakdown of the of the rotation symmetry 
may induce optical and mechanical properties. The flow of a liquid crustal around a particle 
not only depends on its shape and the viscosity coefficients but also on the direction of the 
molecules \cite{tur}. The estimates of particle mobility in the liquid crystal, which were 
derived from those of viscous friction for a size Stokes sphere, give much not right values 
compared with relevant experimental date.

One the possible reasons for such discrepancy may be an increase of the effective mass of 
an particles due formation of the deformation coat around it which moving together with the 
particle. This formation elastic deformation coat troth distribution director field around 
particle increase the friction drag on particle moving in a liquid crystal. In other case
the particle which is foreign in liquid crystal interacting with another similar particle 
\cite{po1}-\cite{lev}. The physical mechanism of this interaction is that the particle 
distorts the director distribution that can occupy a region much greater than its dimensions 
and thus provide an effective interaction with another similar particle via mediation of the 
elastic field deformation. The role this interaction between the colloidal particles is two 
ford. First of all, it enters the particle expression for the viscous-elastic response function. 
These macroscopic expression are ensemble averages of phase function, among which are the 
elastic interaction function. These long-range interaction function are responsible for the 
strong divergence of the viscous elastic response function for colloids as compared to states 
of the all systems. Second, the assemble average that represents the viscous elastic response 
function must be evaluated with respect to the shear rate distorted pair-correlation function. 
The shear rate dependence of this probability density function is the result of an interplay 
between equilibrium restoring forces and shear forces. The anomalous contribution is that part 
of the viscous elastic response function that diverges at the critical point due to the development 
of long-rang correlation induced by particle interaction truth field director deformation. In 
this time, many investigations have been carried out of the rheological properties of colloidal 
liquid crystal suspension \cite{te3}-\cite{st3}. Therefore, a comparison between our results and 
those would only be possible for the Newtonian viscosity, except for the essential difference in 
the interaction between particles and liquid crystal, and particles interaction truth deformation 
elastic field distribution director in the liquid crystal. A macroscopic evaluation of the viscous 
elastic consists of the steps : a) the first step realization formation mesophase with formation 
of a deformation coat ''solvate shell'' around usual particle; b) the second step consist in 
description influence in the particle interaction induced by elastic deformation the director field; 
c) the last step consist the description the probability phase transition in system particle wish 
accompanied the new structure formation, which observed experimentally. The approach proposed in 
this article makes it possible describe the collective diffusion in liquid crystal colloid by all 
area the determination temperature and concentration the matter. We use concept of the effective 
mass and friction drag of colloidal particle which moving in liquid crystal. As shown \cite{tur}, 
along with an particle traveling in liquid crystal, there also moves a director deformation field 
linked with extremely weak anchoring on the surface.

The first one deals with spherical particle that have a strong anchoring strength on the surface 
\cite{lupe}, \cite{te1}. It creates topological defect in the vicinity of the particle which are 
necessary to satisfy the topological global boundary conditions. A particle with strong planar 
anchoring creates a pair of topological defects, known as boojums . A particle  with strong 
homeotropic boundary conditions, on the other hand, are to create an equatorial disclination 
ring or a hyperbolic hedgehog as a companion for the radial hedgehog on the surface of the particle. 
Using the variational techniques and an electrostatic analogy, Lubensky et al. \cite{lupe} obtained 
an approximate director distribution near the particle with normal boundary conditions, as well as 
the long range pair interaction potential between the particles. 

The second approach was proposed in \cite{lev}, where the authors have examined the case of weak anchoring strength for particles
of a general shape. They have found analytically  the pair interaction potential, taking into account the different Frank
constants and have expressed the potential in terms of a tensor expressing of the shape of the particle. In the article \cite{lev1} 
was argue that the long range interaction potential between particles in liquid crystal is determined by symmetry breaking of the 
director field in the vicinity of the particles. This symmetry breaking is caused by two reasons: the shape of the particle and 
the anchoring strength. In the case of weak anchoring it is determined primarily by the form of the particle. In the case of 
the strong anchoring, on the other hand, both factors are essential, because the director distribution near the particle in this 
case is determined by topological defects in its vicinity. In order to universally describe all these phenomena, was introduce 
the concept of the deformation coat around the particle. The deformation coat embraces all the accompanying topological defects, 
white it has the same symmetry as the resultant director field near the particle. The director distribution outside the coat 
undergoes only smooth variations and does not contain any topological defects. This colloidal particles may also be regarded as 
micro particle surrounded by a ''solvate shell'' provided the interaction between such a particle and the molecules is much more 
intense than the intermolecular interaction responsible for the liquid crystal formation. The solvable formation may be regarded 
as a macro-particle, thus its interaction with another similar formation can be described in term of the director field deformation. 
Taking into account the director distribution around separate particle, one can find the change of orientation of the director 
induced by two particles, and determine the change of deformation energy at the approach of those with separation of the energy 
component corresponding to interaction between those.

In the case then the interaction of the individual particle with the director field deformation produced by the other particles is
determined by the anchoring on the surface of this particle is describe in article \cite{lev}. When the number particles is small 
and the week anchoring on the surface, which realize in our situation the colloidal particles in liquid crystal, the distribution 
the director field is determined in papers \cite{te4}. In next we briefly describe this approach. Strong anchoring directly implies 
that close to the colloidal surfaces significant spatial variations of the director and even defects can appear, meaning that 
assumption of a roughly uniform director $\mathbf{n}=(n_x,n_y,1)$ is not valid anymore throughout the whole liquid crystalline 
volume. In order to explain subsequent speculations we consider first the case of homogeneous liquid crystal with uniform director
$\mathbf{n_{0}}$ and one particle immersed into it. Anchoring of the liquid crystal with the surface of the particle deformations
of the director field around the particle so that director $\mathbf{n(\mathbf{R})}$ varies from point to point. In the one-constant 
approximation the total free energy of the system:
\begin{equation}
F=\frac{K}{2}\int dV \left[(div
\mathbf{n})^{2}+(rot\mathbf{n})^{2}\right]+\oint dS W(\mathbf{\nu
}\cdot \mathbf{n})^{2}\label{fto}
\end{equation}
where $W$ is anchoring strength coefficient, $\nu$ is the unit normal vector, integration $\oint dS$ is carried out on the surface of 
the particle, $K$ is the Frank elastic constant. Far from the particle director field variations are small $\mathbf{n(\mathbf{R})}\approx(n_{x},n_{y},1)$, $|n_{\mu}|\ll 1$
($\mu=x,y$) and bulk free energy has the form:
\begin{equation}
F_{b,linear}=\frac{K}{2}\int dV \left\{(\nabla n_{x})^{2}+(\nabla
n_{y})^{2}\right\}\label{flin}
\end{equation}
which brings Euler-Lagrange equation of Laplace type:
\begin{equation}
\Delta n_{\mu}=0 \label{nmu}
\end{equation}
At large distances $R$ in general case it can be expanded in multiples,
\begin{equation}
n_{\mu}(\mathbf{R})=\frac{q_{\mu}}{R}+\frac{\textbf{p}_{\mu}\textbf{u}}{R^{2}}+3\frac{\textbf{u}:\hat{Q}_{\mu}:\textbf{u}}{R^{3}} \label{nmufar}
\end{equation}
with $u_{\alpha}=R_{\alpha}/R$; $\textbf{p}_{\mu}\textbf{u}=p_{\mu}^{\alpha}u_{\alpha}$, $\textbf{u}:\hat{Q}_{\mu}:\textbf{u}=Q_{\mu}^{\alpha\beta
}u_{\alpha}u_{\beta}$, and $\mu=(x,y)$. This is the most general expression for the director field. We note that the multiple expression does 
not depend on the anchoring strength. It is valid on far distances for any anchoring, weak and strong, without topological defects or with them. 
Of course in order to find multiple coefficients we need to solve the problem in the near nonlinear area either with computer simulation or Ansatz
functions. Let's imagine that we have found all multiple coefficients for the particular particle (for instance with computer simulation). 
After this presentation we can use both approach to the description energy interaction independence from value anchoring strength on the
surface particle. For particles with strong anchoring it is the far-region, because of strong director deformations in the near-region. But for
particles with weak anchoring distortions are small elsewhere and the multiple expansion is applicable in the near-region too. The symmetry of 
the coat is equivalent to the broken symmetry of the director in the vicinity.  

A way how to avoid strong deformations of the director field and incorporate them into an analytical description is to introduce a ``coat region'' 
\cite{lev1}. This region acts as an effective colloid which incorporates all strong deformations of the director field around the real particle 
and has the same symmetry as the director distribution. In article \cite{cher} the general paradigma of the elastic interaction  between colloidal 
particles in nematic liquid crystal was proposed according to each every particle with strong anchoring and radius $R$ has three zones around itself. 
The ﬁrst zone for $r < 1.3R$ is the zone of topological defects, and for determination of the distribution director field must use the nonlinear
equation; the second zone at the approximate distance range $1.3R < r <4R$  is the zone where crossover from topological defects to the main multiple 
moment takes place. The last third zone - is the zone of the main multiple moment, where higher order terms can neglect. It is possible simple 
explanation of these presentation. In order to understanding and by definition size of this zones we can propose next way.

The task of finding a director distribution around the spherical particle we must minimization of Frank energy and take into account the
boundary condition. This task was solved in article \cite{te2}. In spherical coordinate we can take  $\vec{n}( \vec{r})=(\sin \beta
\sin \gamma ,\sin \beta \cos \gamma ,\cos \beta )$, where $\beta$-polar and  $\gamma $ - azimuthal angle. In this representation exist azimuthal 
symmetry relatively $z$. The difference equation in one constant approximation takes the form :
\begin{equation}
\nabla \beta (r)-\frac{\sin 2\beta (r)}{2r^{2}\sin ^{2}\theta }=0
\end{equation}
and the boundary condition on the surface the particle 
\begin{equation}
\left(\frac{\partial\beta}{\partial r}+\frac{\beta}{r}_{r=R}=-\frac{W}{2K}sin\theta\right)
\end{equation}
have the general solution $\beta=\sum_{k}\frac{C_{k}}{r^{k+1}}P_{k}^{1}(cos\theta)$ where $P_{k}^{1}(cos\theta)$ associated with Legendre 
polynomial. The boundary condition selects the solution , which in the case week anchoring $(R W/K)\leq 1$ takes the form \cite{te1}: 
\begin{equation}
\beta (r)=(RW/4K)\bigskip R^{3}\sin 2\theta /r^{3}
\end{equation}
where $W$ - anchoring energy and $R$ - radius of spherical particle. In the case small deformation of director field $(R W/K)<4$. 
This relation determine the condition of week anchoring. In the case strong anchoring $(R_{0}W/K)\geq 1$ around the spherical particle 
appear disclination ring with radius $a$. This solution also was obtain in the article \cite{te2} and take the form 
\begin{equation}
\beta (r)=\theta -\frac{1}{2}\arctan \frac{\sin 2\theta }{\cos2\theta +( \frac{a}{r})^{3}}
\end{equation}
Inside the the disclination ring the solution of distribution of director field take the form $\beta=\left(\frac{a}{r}\right)^{3}\sin 2\theta$ 
where $a$ radius of ring, which can be obtain from minimum of free energy. Can estimate the radius of ring in expression $a=\frac{5}{4}R$. 
In order to that disclination ring disappear around the spherical particle is necessary such anchoring as this solution must be solution 
as in case week anchoring. Therefore was content ourselves with the quality estimate with crossover value $W^{\ast}\sim
\frac{4Ka^{3}}{R^{4}}=\frac{125}{16}\frac{K}{R}$. In the case the disclination ring not appear around the spherical particle. In this 
approach was obtained the free energy which introduce the one particle in liquid crystal \cite{te1}, \cite{lupe}. In the case week anchoring 
$F\approx \frac{W^{2}R^{3}}{5K}$ and in the case strong anchoring  $F\approx 13KR$. The uniform director distribution far from the particle has 
zero topological charge and so there should be another topological defect near the particle to compensate the hedgehog in the center. 
Obviously, the director configurations have different symmetry. The non equatorial disclination ring and the pair of radial and hyperbolic 
hedgehogs break mirror symmetry in the horizontal plane, while equatorial disclination ring (Saturn-ring) retains it.  Authors of \cite{te1} 
have shown by Monte-Carlo simulations, that the configuration with hyperbolic hedgehog has lower energy, than Saturn-ring. 
It has been confirmed in \cite{lupe} with help of the dipole Ansatz , that though the equatorial ring has some metastability, 
its energy is higher, than of the dipole configuration.

We can introduce the deformation coat which include all strong deformation of director field. This deformation area is
new ``immersed particle'' and we can use only self-consistence approach in the case week anchoring and small deformation of
director field inside this new inclusion. We can estimate the size this deformation coat if we take into account the all energy
which introduce in liquid crystal own particle. For the spherical particle with radius $R$ in the case strong anchoring free energy
of deformation can present in the form \cite{lupe} $F^{strong}\approx \frac{9}{2}\pi K R$. We assume that around the spherical particle 
exist the Saturn-ring disclination. The size of deformation area denote as $R^{\ast}$. As next step we calculate the free energy which 
can introduce the deformation area inside. This energy can obtain if suppose that the inside this deformation area we have the case week 
anchoring an can to make well use the distribution director field as in case week anchoring. This energy was obtain in the article\cite{te1}
$F^{week}=\frac{\pi}{15}\frac{W^{2}_{c}R^{\ast 2}}{K}$, where $W_{c}$ critical value of energy anchoring for the case different
size of particle when outside the particle not appear peculiarity in the distribution of director field. This critical value was
obtain also in the article \cite{te1} and is $W_{c}=\frac{4K a^{3}}{R^{4}}$ where $a$ radius of Saturn-rind disclination around spherical 
particle. Inside the deformation area exist the disclination ring . Free energy of this Saturn-ring disclination we can present in the form 
\cite{lupe} $F^{disc}\approx 2\pi K a\left(\frac{3\pi}{4}+1\right)+8\pi K(R^{\ast}-a)$ , where size of Saturn-ring can be present in the 
previous form. For estimation of size of deformation coat we can compare the free energy which real create inclusion in the case strong
anchoring to sums free energy which create deformation coat outside and free energy Saturn-ring disclination inside this coat. From relation 
$F^{week}+F^{disc}=F^{strong}$ the size of deformation area take the form $R^{\ast}\approx 2R$. This estimation was made in the case Saturn-ring 
disclination inside the deformation coat but this approach can be use in the case different defect which can appear inside this coat. In the 
case dipole configuration in distribution director field inside this are coerced estimation is correct to, but the deformation coat will have 
the asymmetric form. The size of deformation coat in all case have the same approximate size.

When the determined the size area the elastic deformation director, we have possibility study friction drag and inertial characteristic particle.
The friction drag on particle moving in nematic liquid crystal is determined as results computer calculation in the article \cite{te4}, \cite{st1}. 
We focus our attention on the effective mass usual particle which moving in liquid crystal. As shown earlier, along with an
particle traveling in liquid crystal, there also moves a director deformation field linked with the surface anchoring on this particle.
Thus, an particle moving inside nematic liquid crystal the kinetic energy $T=mu^{2}/2$ that the deformation coat \cite{lev4},\cite{lev1}. We
assume that the deformation coat adiabatically follows a moving particle; a similar assumption was used by in constructing the polaron
theory. In this connection,it should be noted that the assumption used implies the following restriction on velocity of the particle $u\leq \tilde{
R}t_{0},$ where $t_{0}$ is a time interval during which the director transition to steady state takes place relevant estimates of a spherical
size $\tilde{R}$ of the deformation coat. In the framework of conception of deformation coat, the director distribution around an a particle 
naturally determined as $\textbf{n}=\textbf{n}(t-\int u(t^{\prime })dt^{\prime }).$  Numerical estimation of the effective mass can be obtained 
if one known the director distribution around a moving particle. When take the distribution director in the case week anchoring, which is description 
by \cite {te4} isotropic part effective mass the particle which moving in liquid crystal we arrive at the following as expression 
$m_{eff}=m+\frac{4I}{3}(\frac{W}{4K})^{2}R^{3}$, according to results of the articles \cite{lev4}, where $I$ is the density of the nematic 
liquid crystal moment of the inertia.

In the approach the Stokes-Einstein relation between coefficient diffusion end the mass Brownian particles $D=kT/6\pi \mu \tilde{R}$ \ \ ($\mu $ is the
viscosity coefficient) we could be experimental date decrease the coefficient diffusion on the second region decreasing temperature. In this
region take the phase transition in the liquid crystal and formation the deformation coat around usual particle. This phenomena is possibility 
decreasing the mobility particles which motivation the increasing friction drag. Thus, it becomes possible to attribute a decrease of the particle 
mobility, observed under the transformation into the liquid crystal, to the deformation coating effect and, consequently, to the increasing effective 
mass of a particle.

To describe the peculiarities of the particle system behavior in the liquid crystal implies taking into account interaction via the director elastic
field. We have already shown that a foreign particle produces liquid crystal distortion in a region much greater than the particle dimensions and
thus leads to an effective interaction with another similar particle via the director field deformation.  In this sense, the interaction of the 
spherical particle also is associated with the director elastic field deformation. The particle dispersed in the liquid crystal cause long ranged
deformation of the director field. The self-consistent approach provides a possibility to avoid the above difficulties, so we have managed to find the
energy of the inter-particle interaction of particles introduced in the liquid crystal. Having found the inter-particle interaction energy,
we can study the thermodynamic behavior of an aggregate of such particles and describe the condition for the creation of new structure \cite{lev2,lev1}.
The character and intensity of the inter-particle interaction in the system of foreign particles in liquid crystal can be such that a
temperature and concentration phase transition in the system and produce a spatially inhomogeneous distribution the particles \cite{lo}. The is
first-order face transition when the external field present itself the topological defect in nematic ordering. The type of interaction between the 
colloidal particles is sufficiently long ranged to lead to a strong critical divergence of the shear viscosity. It is therefore interesting to study
the full temperature dependence of the shear viscosity of colloidal system near the critical point with transition in new phase and peculiarity 
when this transition accompanied the special inhomogeneous distribution in system particles. The viscous elastic response function will turn out to 
be equal to two distinct additive contribution : an anomalous and a background contribution. The anomalous contribution is that part of the viscous 
elastic response function that diverges at the critical point due to the development of long-range correlation. The background contribution must 
be subtracted from experimental viscous elastic response function to obtain their observer anomalous contribution, which may then be compared to 
theoretical prediction. Therefore a comparison between our results and those would only be possible for Newtonian viscosity, except for the essential 
difference in the inter-particle potential. One of the pictorial structures is inadequate property to describe the substance. The distribution of 
viscous elastic properties in the system interacting particles is known in papers \cite{oh}-\cite{ver} sufficient detail. We designer must fall 
back on methods of analysis the viscous elastic properties in colloidal suspension .

As you are well aware, furthermore $D_{eff}(\vec{k})$ wave-vector-dependent effective diffusion coefficient equal to \cite{oh}-\cite{ver}
\begin{equation}
D_{eff}=\frac{D}{kT}\{\frac{dG}{d\widetilde{\rho }}+q^{2}S\}  \label{10}
\end{equation}
where $D$ is the usual diffusion coefficient spherical particle which moving in condense matter. The first on the right-hand side of this equation
describes shear flow distortion the interacting Brownian particle, and have representation in the form

\begin{equation}
G=\widetilde{\rho }kT-\frac{2\pi }{3}\widetilde{\rho }^{2}c(1-c)\int
dr^{\prime }r^{\prime 3}\frac{dV(r^{\prime })}{dr^{\prime }}g^{eq}(r^{\prime
})  \label{11}
\end{equation}

where $\widetilde{\rho }$ is concentration of the media, $c$ is friction parameter particles which is foreign in liquid crystal, $V(r^{\prime })$
is the pair interaction energy between particles truth the elastic deformation director field, and $g^{eq}(r^{\prime })$ is the pair correlation
function. The last term in Eq. (8) describes the diffusion limited tendency to restore the equilibrium structure with
\begin{equation}
S=\frac{2\pi }{15}\widetilde{\rho }c\int dr^{\prime }r^{\prime 5}\frac{
dV(r^{\prime })}{dr^{\prime }}\{g^{eq}(r^{\prime })+\frac{1}{8}\widetilde{
\rho }(1-c)\frac{dg^{eq}(r^{\prime })}{dr^{\prime }}\}  \label{12}
\end{equation}
Close to the critical point and also close to the of-critical part of the spinodal decomposition, which accompanied the first-order phase
transition,, where $\frac{D}{kT}\frac{dG}{d\widetilde{\rho }}$ is small, the effective diffusion coefficient is small for small wave vectors, 
a phenomenon that is commonly referred to as critical slowing down.

As we will be show, the general solution of diffusion coefficient can be expressed in the term of explicit expressions for structure factor for two
more simple cases: the structure factor in the small inter-particle interaction truth the deformation director field of shear flow and under
stationary shear flow in the present the inter-particle interaction with phase transition in spatial inhomogeneous distribution this particles.
The above equation relate to the mean-field behavior of the structure factor. This is due to transmutation the present relation to form;
\begin{eqnarray}
&& G=\widetilde{\rho }kT+\frac{2\pi }{3}\widetilde{\rho }^{2}c(1-c)\int
dr^{\prime }r^{\prime 2}V(r^{\prime })g^{eq}(r^{\prime })\nonumber\\
&&-\frac{2\pi }{3} \widetilde{\rho }^{2}c(1-c)\int dr^{\prime }r^{\prime 3}V(r^{\prime })\frac{
dg(r^{\prime })}{dr^{\prime }}) \label{13}
\end{eqnarray}
The stiffness of the bond is lessened resulting in a lowering of dependence from temperature the effective diffusion coefficient in follows form:
\begin{equation}
D_{eff}=D\{1-\frac{T_{c}-a}{T})  \label{14}
\end{equation}
where $a=\frac{2\pi }{3}\widetilde{\rho }^{2}c(1-c)\int dr^{\prime }r^{\prime 3}V(r^{\prime })\frac{dg(r^{\prime })}{dr^{\prime }}$, and
\begin{equation}
T_{c}=-\frac{2\pi }{3}\widetilde{\rho }^{2}c(1-c)\int dr^{\prime }r^{\prime
2}V(r^{\prime })g^{eq}(r^{\prime }) 
\end{equation}
is the critical temperature first-order phase transition,with accompaniment the formation spatial inhomogeneous structure in the distribution in system
particles. The relevance of the correlation length is that it measures the range over which colloidal particles in the unsorted system are correlated. 
Since $\frac{dG}{d\widetilde{\rho }}\rightarrow 0$ on approach of the critical point and also on approach of the off-critical part of the spinodal 
decomposition the correlation length $\xi =\sqrt{S/\frac{dG}{d \widetilde{\rho }}}$ diverges. This means that at critical point each colloidal particle 
in system is correlated with all other colloidal particles. One may imagine that it will take on finite force to break up these many correlation in 
order to make the system flow, which means that the viscosity diverges on approach of the spinodal. The effective coefficient diffusion have the 
dependence of correlation length in follow form $D_{eff}=D\xi ^{-2}$ illustrate the decries the time relaxation from the needier the temperature 
the firs-order phase transition.

In order case, when the has to be taken into account all physical processes, which motivation the structure transition in our matter for this
fact behavior the effective diffusion coefficient maybe describe the follow physical picture: When the formation the liquid crustal, every particles
dresses in to elastic deformation coat the inhomogeneous distribution director field around the particle. This fact provide to increase
effective mass every particle an increase the drag force motion usual particle and decrease the effective coefficient diffusion. When the liquid
crystal are formation, generation the interaction between particle by induced the director field deformation. 
This repulsive interaction truth the elastic field deformation provide to increase the collective effective diffusion coefficient. When the interaction 
in system particles motivate the firs-order phase transition which accompanied the inhomogeneous the distribution particle the fact are the decrease 
the effective diffusion coefficient. This fact are attributable to realization long range correlation in system particle, which is foreign in liquid crystal.
Based on this physical picture and an experimental date the complete behavior the collective diffusion coefficient in our case is rendered possible 
by the addition of the approximation formula:
\begin{equation}
D_{eff}=D_{0}\left[ 1-b(T-T_{in})^{2}\right] \{1-\frac{T_{c}-a}{T})
\label{16}
\end{equation}
where $T_{in}$ is temperature the phase transition our media in liquid crystal state and $b$-coefficient which determined dependence the effective
mass(7) or friction drag the moving particles which is foreign in liquid crustal. This relation describe the all region dependence the effective
collective diffusion coefficient from the temperature and concentration particles in our media. The basic pattern of fracture can be produced at 
will by establishing appropriate experimental conditions.
As conclusion, we can note about the fact peculiarity behavior the effective diffusion coefficient which take into account the dresses every particles
in to elastic deformation coat. This fact provide to increase effective mass every particle an increase the drag force motion usual particle and 
decrease the effective coefficient diffusion. When the liquid crystal are formation, generation the interaction between particle by induced the director 
field deformation. The interaction truth the elastic field deformation provide to increase the collective effective diffusion coefficient. When the interaction 
in system particles motivate the firs-order phase transition the fact are the decrease the effective diffusion coefficient. This fact are attributable 
to realization long range correlation in system particle. Based on this physical picture is possible experimental observe the complete behavior the 
usual and collective diffusion process.

\textbf{Acknowledgement:} this research was partially funded by the State Fund for Fundamental Research of Ukraine (Project №~F~76/84).

\end{document}